# Electronic phase transitions and superconductivity in ferroelectric $Sn_2P_2Se_6$ under pressure

He Zhang[1, 2#], Wei Zhong[3#], Xiaohui Yu[1,2,4*], Binbin Yue[3]*, Fang Hong[1,2,4]*

[1] *Beijing National Laboratory for Condensed Matter Physics, Institute of Physics, Chinese Academy of Sciences, Beijing 100190, China*

[2] *School of Physical Sciences, University of Chinese Academy of Sciences, Beijing 100190, China*

[3] *Center for High Pressure Science & Technology Advanced Research, 10 East Xibeiwang Road, Haidian, Beijing 100094, China*

[4] *Songshan Lake Materials Laboratory, Dongguan, Guangdong 523808, China*

[#] These authors have equal contribution.

[*] Email: yuxh@iphy.ac.cn, yuebb@hpstar.ac.cn; hongfang@iphy.ac.cn

**Abstract**

Since there is both strong electron-phonon coupling during a ferroelectric/FE transition and superconducting/SC transition, it has been an important topic to explore superconductivity from the FE instability. $Sn_2P_2Se_6$ arouses broad attention due to its unique FE properties. Here, we reported the electronic phase transitions and superconductivity in this compound based on high-pressure electrical transport measurement, optical absorption spectroscopy and Raman based structural analysis. Upon compression, the conductivity of $Sn_2P_2Se_6$ was elevated monotonously, an electronic phase transition occurred near 5.4 GPa, revealed by optical absorption spectroscopy, and the insulating state is estimated to be fully suppressed near 15 GPa. Then, it started to show the signature of superconductivity near 15.3 GPa. The zero-resistance state was presented from 19.4 GPa, and the superconductivity was enhanced with pressure continuously. The magnetic field effect further confirmed the SC behavior and this compound had a $T_c$ of 5.4 K at 41.8 GPa with a zero temperature upper critical field of 6.55 T. The Raman spectra confirmed the structural origin of the electronic transition near 5.4 GPa, which should due to the transition from the paraelectric phase to the incommensurate phase, and suggested a possible first-order phase transition when the sample underwent the semiconductor-metal transition near 15 GPa. This work demonstrates the versatile physical properties in ferroelectrics and inspires the further investigation on the correlation between FE instability and SC in $M_2P_2X_6$ family.

**Introduction**

Ferroelectric (FE) materials are of great importance to fundamental research and realistic application, and the related study has been last more than 100 years [1]. The most common ferroelctrics are in form of oxides, such as $BaTiO_3$, $PbTiO_3$, lead-free niobates, and so on. With the development of advanced characterization techniques and theoretical modeling, the reception on ferroelectrics has been extended widely. Now, ferroelectricity can be treated as a formalism of the quantum mechanical Berry-phase [1]. In some magnetic compounds, ferroelectricity can be driven by unique magnetic orderings, such as spiral spin ordering, and it can even coexist with ferromagnetic ordering [2]. Beyond the oxide FEs, various new types of FE materials are reported, from metal chalcogenides [3,4], polymers [5,6], metal–organic frameworks [7], to sliding two-dimensional bilayers [8-10]. Benefiting from the switchable polarization, ferroelectrics have been used as sensors, energy transformation devices, FE random access memories, magnetoelectrics *et al.*

In most cases, FE states are presented in materials with a band gap since free electrons can screen the long-range Coulomb forces and doesn't favor the polar state [11]. By tuning the chemical composite or applying pressure on FE materials, PE-FE transition can be well controlled, FE can be even suppressed, which may result in a quantum paraelectric phase or FE quantum critical point [12-14]. Since there are both strong electron-phonon coupling during the PE-FE transition and in BCS superconductors, superconductivity could emerge due to the instability of FE. Typical example is the doped $SrTiO_3$, near its FE quantum critical point, the superconductivity is found to be mediated by the longitudinal hybrid-polar-modes recently based on high-pressure experiments [15], and it can be enhanced by pressure-driven plastic deformation [16], which enhances the soft polar fluctuations, though previous studies claim an unconventional superconductivity [17]. The electron doping can also induce the instability of FE soft mode in $LaOBiS_2$ and then superconductivity emerges together with a distorted charge density wave [18]. Therefore, it will be of great interests to study the underlying exotic physics in FEs by external stimulation, such as doping and pressure.

$Sn_2P_2X_6$ (X= S, Se) compounds were reported to be ferroelectric [4,19,20]. For $Sn_2P_2S_6$, it is a room temperature FE materials with Tc at ~337 K, while the Tc reduces to ~193 K in $Sn_2P_2Se_6$. Recently, $Sn_2P_2S_6$ is found to be a superconductor under pressure when it transforms from the paraelectric phase to a chiral $P2_1$ phase, and the superconductivity is expected to be related to the flat band in the chiral phase[21]. Though $Sn_2P_2Se_6$ looks similar with its sister compound $Sn_2P_2S_6$, as $Sn_2P_2S_6$ and $Sn_2P_2Se_6$ share the same paraelectric ($P2_1/c$) and ferroelectric ($Pc$) structures, their paraelectric-ferroelectric phase transition processes are different actually, as there is intermediate incommensurate phase in $Sn_2P_2Se_6$ locating between the high temperature paraelectric phase and low temperature ferroelectric phase [20]. The incommensurate (IC) phase is an iso-symmetric structure of the paraelectric phase. For the PE-FE transition in $Sn_2P_2S_6$, it is a second-order transition, but it is a first-order transition in $Sn_2P_2Se_6$ during the IC-FE transition [22]. Considering these differences, we studied the electronic behavior of $Sn_2P_2Se_6$ by means of high-pressure transport, optical absorption and Raman spectroscopy. $Sn_2P_2Se_6$ underwent an electronic phase transition near 5.4 GPa with a big drop of optical band gap, and it is probably due to the paraelectric-incommensurate phase transition, signaled by the two new Raman modes at low wavenumber region. The optical band gap was estimated to close near 15 GPa, and then superconductivity emerged near 15.3 GPa, which was enhanced with pressure monotonously, and the SC critical temperature $T_c$ increased from ~2.2 K at 15.3 GPa to 5.4 K at 41.8 GPa. Meanwhile, Raman signal was still detectable between ~15 and ~30 GPa, suggesting a possible first-order phase transition when the superconductivity was presented. And this assumption could be supported by the different pressure dependent behavior of Tc below and above 30 GPa. Considering the similar effect on the lattice structure between temperatures cooling and pressure, the phase transition under pressure may follow the sequence as observed during the cooling process at ambient pressure. Then, the high-pressure phase could be in a FE state or near the FE quantum critical point. Further detailed structural analysis based on synchrotron x-ray diffraction is still required. This work has demonstrates the abundant phase transitions in ferroelectric $Sn_2P_2Se_6$ under pressure, and is beneficial for the future study on the correlation between FE instability and superconductivity in this kind of FE materials.

**Experiment**

*High-pressure electric transport measurements*

The four-probe electrical resistance measurement was carried out in a van der Pauw geometry under high pressure up to ~43.5 GPa, and the experiment was conducted in a commercial cryostat from 1.7 K to 300 K by a Keithley 6221 current source and a 2182A nanovoltmeter. Two opposing anvils with 300 μm culets were placed in a BeCu alloy diamond anvil cell (DAC) to generate high pressure. A thin single crystal sample was loaded into the sample chamber in a rhenium gasket with c-BN insulating layer, and a ruby ball was loaded to serve as an internal pressure standard. KBr was used as the pressure medium. To reduce the pressure gradient of the sample environment, a tiny sample with size of ~30 μm (W) *~40 μm (L)*~5 μm (H) was cut from original bulk crystal and placed in the center of sample chamber.

*High-pressure spectroscopy measurements*

The high-pressure Raman spectra were collected using a Renishaw Micro-Raman spectroscopy system equipped with a second-harmonic Nd:YAG laser (operating at 532 nm). The laser power was maintained at relatively low power level to avoid overheating the sample during measurement. In-situ high-pressure UV-VIS-NIR absorption spectroscopy was performed on a home-designed spectroscopy system (Ideaoptics, Shanghai, China). In all spectroscopy measurements, type IIa diamond anvils with low fluorescent signal were used and KBr was used as pressure medium. Pressure was determined from the shift of $R_1$-$R_2$ ruby fluorescence lines [23].

**Results and discussion**

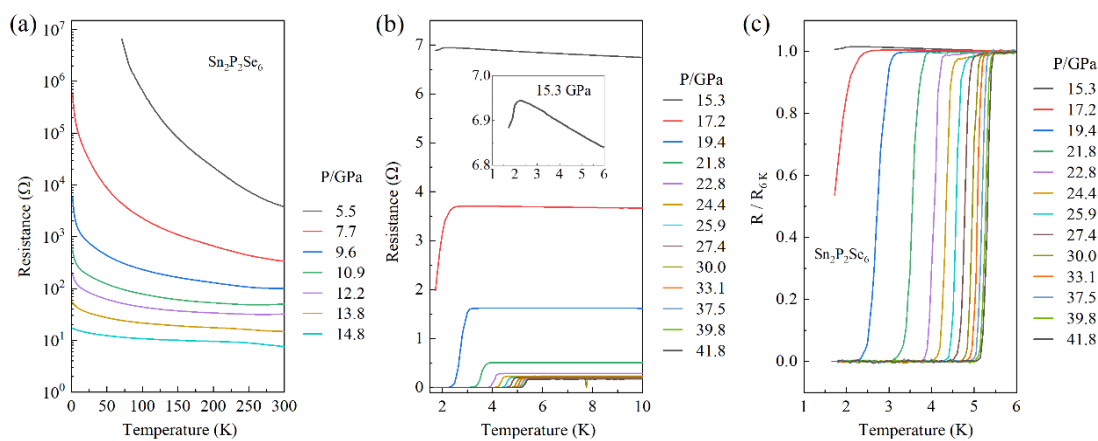

**Fig.1 The electrical transport properties of Sn$_2$P$_2$Se$_6$ under pressure.** (a) The R-T curves from 1.8 K to 300 K under low pressure range from 5.5 to 14.8 GPa, showing a semiconducting behavior. (b) The R-T curves from 15.3 to 41.8 GPa, showing superconducting/SC transitions. (c) The normalized R-T curves at 6 K near the SC transition.

The starting materials, Sn$_2$P$_2$Se$_6$ is a semiconductor, and the initial resistance of the sample is too large and out of the detection limit of the experimental set-up. By adding some pressure, we were able to collect R-T curve staring from ~5.5 GPa. As shown in Fig. 1(a), the resistance of the sample at room temperature and 5.5 GPa is on scale of several thousand Ohm, and it increases by ~ 3 orders of magnitude at ~75 K, indicating an insulating ground state. However, the conductivity will increase upon further compression. The semiconducting behavior persists until 14.8 GPa, above which a small drop of resistance occurs below 2.2 K at 15.3 GPa, as seen in Fig. 1(b). This is a signature of superconducting transition, and such a resistance drop behavior is enhanced at higher pressure, and zero-resistance state is observed at 19.4 GPa. The SC transition temperatures ($T_c$) shift to higher temperature as pressure increases. It reaches 5.4 K at 41.8 GPa, the highest pressure in current work. To display the trend of SC transitions clearly, the R-T curves are normalized and the results are presented in Fig. 1(c). It is noted that the $T_c$ increases quickly at low pressure range and seems to saturate near 40 GPa. Previous room temperature resistance measurement on $Sn_2P_2Se_6$ single crystal also showed a significant change under high pressure, the initial resistance of their sample was on scale of ~$10^{10}$ Ω, and it reduced to only several ohm above 15 GPa[24]. Such a change is almost consistent with our work.

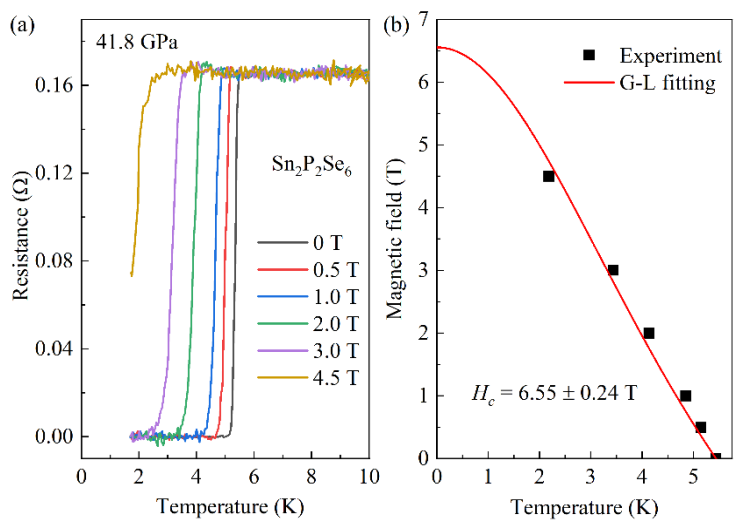

**Fig.2 The magnetic field effect on the superconducting transition in $Sn_2P_2Se_6$ at 41.8 GPa.** (a) The R-T curves at various fields from 0 to 4.5 T. (b) The relation between upper critical field (H) and temperature (T), and the G-L fitting.

To further verify the SC behavior, magnetic field effect has been studied. The SC transition regions under various fields are given in Fig. 2(a). As the magnetic field increases, the superconductivity is gradually suppressed, signaling by the shift of $T_c$ toward lower temperature, and the zero-resistance state is even not observed at 4.5 T above 1.8 K. After extracting the $T_c$ from the R-T curves, we can plot a relation between the $T_c$ and upper critical magnetic field $\mu_0H_{c2}(T)$, as seen in Fig. 2(b). To obtain the upper crucial magnetic field at zero temperature $\mu_0H_{c2}(0)$, the relation is fitted by G-L equation [25] and it gives a $\mu_0H_{c2}(0)$ of ~6.55 T, which is lower than the Bardeen-Copper-Schrieffer (BCS) weak-coupling Pauli paramagnetic limit of $\mu_0H_p = 1.84T_c \approx 9.9$T for a $T_c$ = 5.4 K, suggesting the absence of Pauli pair breaking [26].

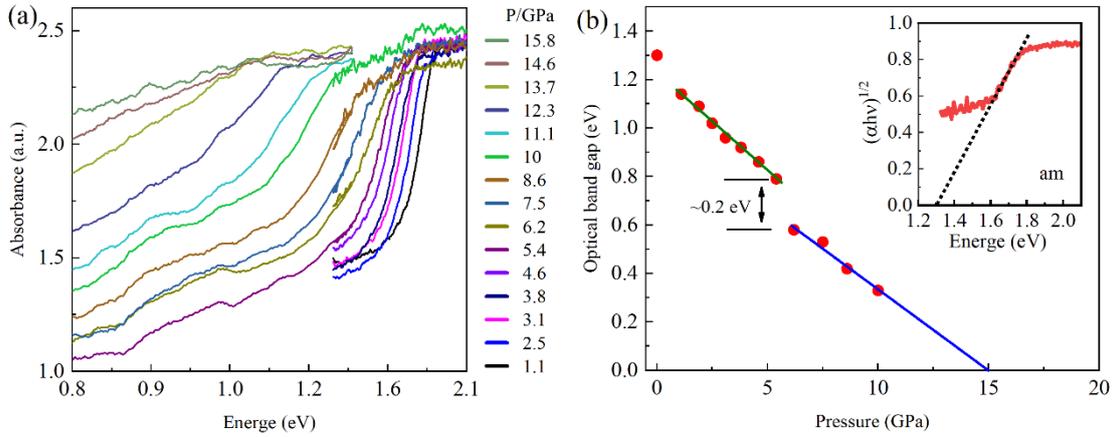

**Fig.3 The optical absorption spectra of $Sn_2P_2Se_6$ single crystal under pressure.** (a) The UV-VIS-NIR absorption spectra of $Sn_2P_2Se_6$ single crystal from 1.1 to 15.8 GPa (the x-axis values are transferred directly from wavelength to energy, and 0.8 eV or 2.1 eV is corresponding to a wavelength of 1600 nm or 600 nm), (b) the relation between fitted optical band gap and pressure, inset: the indirect band gap fitting at ambient condition (am) by using a Tauc relationship.

To understand the electronic behavior before the metallization, we employed UV-VIS-NIR spectroscopy to determine the optical absorption edge and extracted the optical band gap. Since the sample is not a good conductor and shows obvious insulating behavior, the electrical transport measurement cannot give more information about the electronic structure. Here, Fig. 3(a) displayed the optical absorption spectra of $Sn_2P_2Se_6$ single crystal under pressure. Near ambient condition, $Sn_2P_2Se_6$ single crystal has a clear absorption edge between 1.6 and 2.1 eV. By fitting the data with a Tauc relationship, the optical band gap is ~1.3 eV for the ambient sample, as seen in the inset of Fig. 3(b). Such an initial indirect band gap is consistent with the theoretical calculation[24]. Upon compression, the absorption edge moved towards lower energy and became broader, and it was hard to identify the edge above 12.3 GPa. The band gaps at different pressures are presented in Fig. 3(b). As pressure increases, the band gaps show a monotonous decreasing trend. At 1.1 GPa, the band gap reduces to 1.14 eV from original 1.3 eV. After that, the band gap vs. pressure displays a negative linear relationship until 5.4 GPa, above which the band gap experiences a significant drop, and the band gap declines to 0.33 eV at 10 GPa. Though the band gap is hard to determine from the broader absorption edge at high pressure range, we can still make an estimation when the sample could become a metal by extrapolating the band gap-pressure curve in a linear way. The critical pressure for the band gap closing is about 15 GPa, and this value is matched well with the transport measurement. It is noted that the sudden band gap drop near 5.4 GPa signals an electronic phase transition and a possible structural phase transition, which will be discussed later.

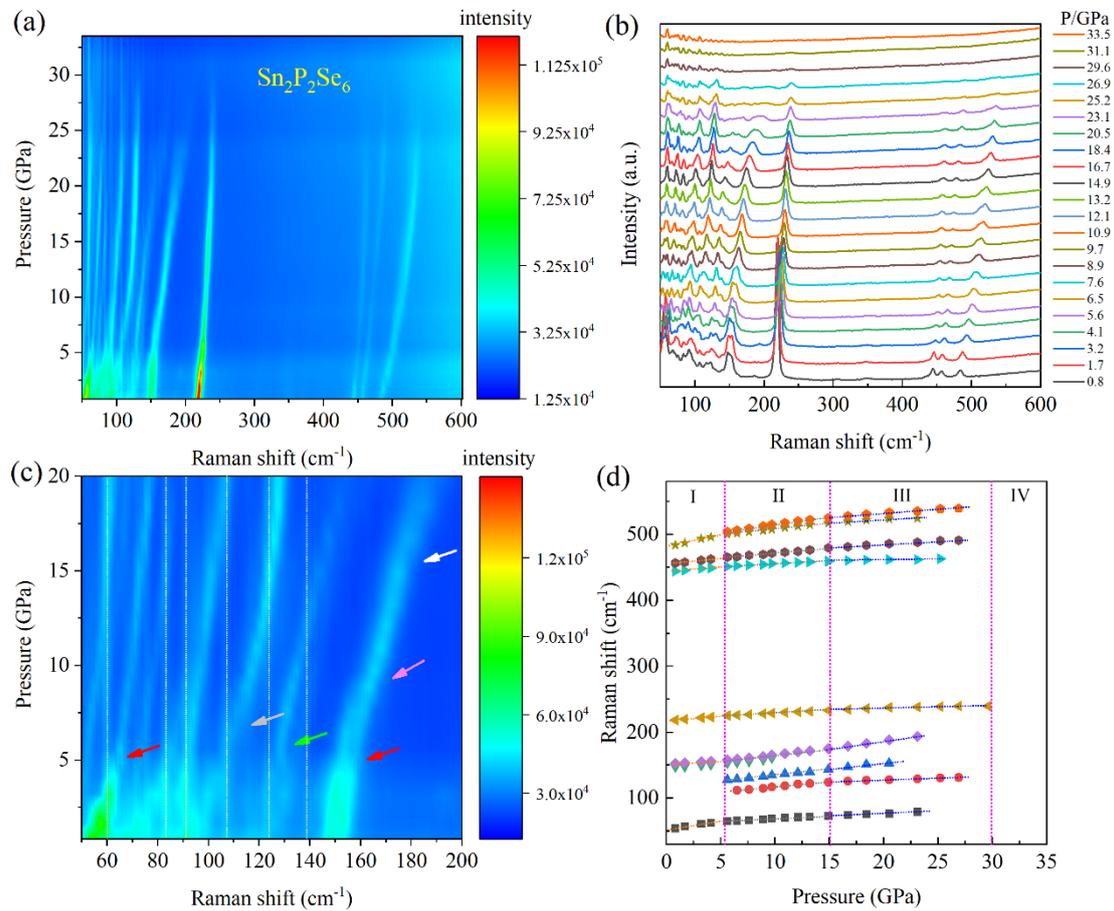

**Fig.4 The Raman spectra evolution with pressure in $Sn_2P_2Se_6$ single crystal.** (a-b) The 2D contour plot and linear plot of the Raman spectra up to 33.5 GPa, respectively. (c) The zoom-in region of Raman at low wavenumber region up to 20 GPa, vertical dash lines are positions of background noisy peaks which don't change with pressure. Arrows marker the changes of Raman

spectra. (d) The relation been various Raman modes and pressure, four regions are proposed for guidance.

To reveal possible structural phase transition, in situ high pressure Raman spectroscopy is employed and the results are displayed in Fig. 4. Near ambient condition (0.8 GPa), there are several relatively strong Raman vibration modes as seen in Fig. 4(a-b), locating at ~150 cm$^{-1}$ (147 and 153 cm$^{-1}$, $A_g$), ~218 cm$^{-1}$ (the strongest mode, $A_g$), ~444 cm$^{-1}$ ($A_g/B_g$), ~456 cm$^{-1}$($A_g/B_g$), and ~483 cm$^{-1}$($A_g/B_g$), which is well consistent with previous works [24]. Upon compression, all of these modes move towards higher frequency, though they show different pressure dependences to some extent, as seen in Fig. 4(a). At the first glance, there is no clear structural phase transitions up to ~30 GPa, since all above main Raman modes are persisted until ~30 GPa. However, Raman peaks are suppressed above ~30 GPa, signaling a complete metallization. The critical pressure for the complete metallization is matched well with previous work [24], in which the Raman peaks disappeared above ~29 GPa. More detailed information can be obtained by zooming in the region of low wavenumber region below 200 cm$^{-1}$. As seen in Fig. 4(b-c), there are actually some changes at low pressure, which are not reported previously [24]. The noisy peaks are marked in Fig. 4(c), and these peaks come from the air and don't change with pressure (the background signal can aslo be well defined by the Raman spectrum (below ~150 cm$^{-1}$) at 33.5 GPa, as seen in Fig. 4(b)). In Fig. 4(c), one clear feature is the intensity drop near ~60 cm$^{-1}$ and ~150 cm$^{-1}$ near 5 GPa. A new peak appears above ~5.6 GPa near 128 cm$^{-1}$, as indicated by the green arrow in Fig.4(c), and another new mode near 112 cm$^{-1}$ is also presented above 6.5 GPa, as indicated by the gray arrow. The Raman change near 5.6 GPa is consistent with the optical absorption results, in which a drop of band gap is presented. Hence, there should be a local structural change which results in the electronic phase transition near 5.4 GPa. Furthermore, the pristine doublet peaks near 150 cm$^{-1}$ merge into one peak near 9.7 GPa (indicated by the pink arrow), and this mode shifts to ~175 cm$^{-1}$ at ~15 GPa. However, the intensity of this peak declines quickly above ~15 GPa (indicated by the white arrow) and becomes broader and broader. The intensity drop and peak broadening are also consistent with previous work [24]. We also note that previous work claims a splitting near 150 cm$^{-1}$ above 11.6 GPa [24]. By comparing the data quality, we believe that their statement should not be correct since their Raman signal was cut off near 150 cm$^{-1}$, which cannot distinguish the sample signal very well from the noisy background.

Considering the similar effect between cooling/temperature and pressure (both of them shorten the atomic-atomic distance and shrink the lattice), the phase transition near 5.6 GPa could be a transition to a low-temperature phase of $Sn_2P_2Se_6$. During this transition, beyond the Raman intensity change, the big difference is the present of two new modes near 128.8 cm$^{-1}$ and 112.2 cm$^{-1}$. At ambient pressure, pristine $Sn_2P_2Se_6$ will undergo a phase transition from the paraelectric phase to the incommensurate phase, before transforming to the ferroelectric phase $P_c$. Such a transition is an iso-symmetric transition occurring at 221 K, both of these two phases are in form of a $P2_1/c$ structure, and it is a second-order phase transition, signaled by two new modes locating at 124 cm$^{-1}$ ($A_g$) and 107 cm$^{-1}$ ($A_g/B_g$) [22]. Taking account of the pressure induced Raman shift, the two new mode near 128.8 cm$^{-1}$ and 112.2 cm$^{-1}$ at 5.6 and 6.5 GPa should origin from the 124 cm$^{-1}$ and 107 cm$^{-1}$ in the incommensurate phase at 220 K. Hence, $Sn_2P_2Se_6$ should transform from the paraelectric phase to the incommensurate phase above ~5.6 GPa at room temperature. Meanwhile, we also noted that earlier study showed a suppression of the incommensurate and ferroelectric phase under external pressure [27]. It seems to be conflict with current work. In fact, the difference could stem from the instability of the paraelectric phase itself under higher pressure. According to previous work [27], the phase transition temperature ($T_i$) of the paraelectric-incommensurate transition declines quickly with pressure, it probably reaches 0 K at only ~1.5 GPa by extrapolating the P- $T_i$ relation linearly, and then the sample is in the paraelectric phase from 0 to 300 K. Under higher pressure, the paraelectric phase is not stable any more, on the contrary, the incommensurate phase is perhaps favored by pressure. Since the transport measurement shows the superconductivity occurs starting from 15.3 GPa, while we can still collect clear Raman signal from the sample, such a behavior is generally due to the mixture phases during a first-order structural phase transition [28]. In other words, there could be a wide pressure window from ~15 GPa to ~30 GPa, during which the $Sn_2P_2Se_6$ sample was in a structural phase transition process: one phase is still the incommensurate phase in an insulating/semiconducting state, while the other is a high-pressure phase showing the metallic and superconducting behavior. Such a first-order phase transition may be consistent with the first-order incommensurate-FE phase transition in $Sn_2P_2Se_6$ sample at low temperature and ambient pressure. Then, it is promising that $Sn_2P_2Se_6$ could be in a FE metal state at high pressure when the superconductivity occurs. Of course, the sample may be in a single phase of a semimetal state above ~15 GPa, which allows us observe the metallic transport behavior and Raman signal simultaneously, as in the case of $Bi_2Se_3$ [29,30]. Further synchrotron x-ray diffraction experiment under high pressure is required to solve the detailed lattice structures.

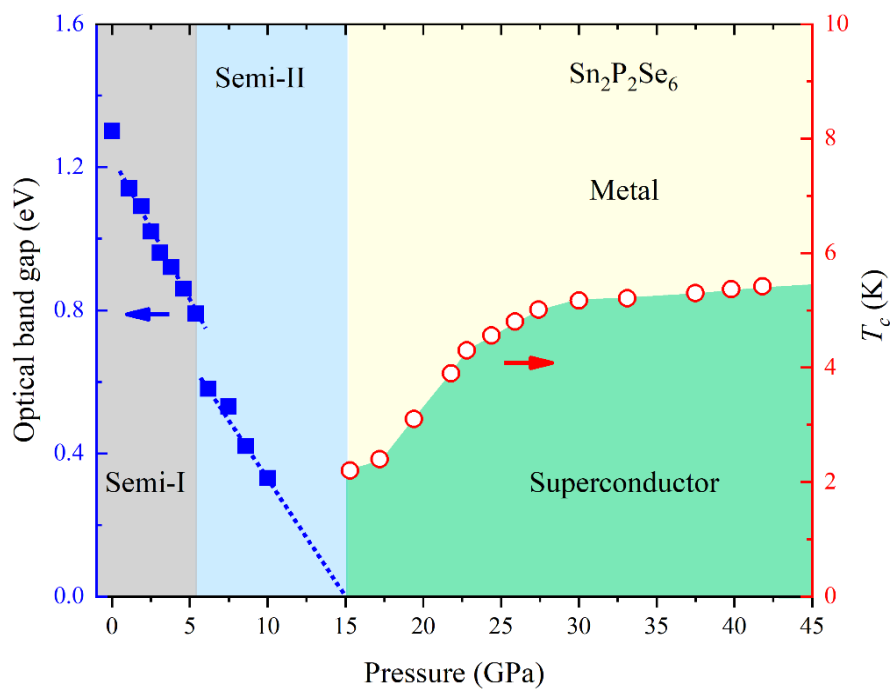

**Fig.5 Proposed electronic phase diagram for $Sn_2P_2Se_6$ under high pressure.** Semi: semiconductor.

Based on the electrical transport measurement, optical absorption and Raman spectra, we propose a brief electronic phase diagram for $Sn_2P_2Se_6$, as displayed in Fig.5. At pressure lower than ~5.4 GPa, the sample is in a Semi-I state and in form of original monoclinic $P2_1/c$ structure. Between ~5.4 and ~15 GPa, the sample probably transforms to the incommensurate phase due to the change of local atomic environment, which is accompanying with an electronic phase transition. Near 15 GPa, the phonon mode near 175 cm$^{-1}$ of the sample started to show the feature of instability, the sample underwent an insulator-metal transition, and show superconducting transition below 2.2 K. Then, the conductivity of sample was further improved and the superconductivity was enhanced, indicated by the quick elevation of $T_c$ values. Between ~15 and ~30 GPa, Raman signal is still detectable though the sample already becomes a superconductor at low temperature, and the sample could experience a first-order structural phase transition in this pressure region. Above 30 GPa, $T_c$ continued to increase with pressure but the increasing rate becomes much slower. This behavior is also consistent with the Raman results, which shows that the sample experienced a complete metallization near 30 GPa, indicated by the disappearance of Raman signal. A detailed structure study is still required to reveal the accurate lattice structures and their evolution under high pressure by using the synchrotron x-ray diffraction.

**Conclusions**

In summary, we reported the superconductivity and electronic behavior in a classic ferroelectric material $Sn_2P_2Se_6$ under pressure by combining electrical transport measurement, optical absorption and Raman spectroscopy. At low pressure (<15.3 GPa), $Sn_2P_2Se_6$ always shows an insulating/semiconducting behavior. Then, superconducting transition starts to appear from 15.3 GPa and it is enhanced upon further compression. The $T_c$ reaches 5.4 K at 41.8 GPa and it is not saturated yet though the pressure dependent increasing rate becomes slower, compared with that below 30 GPa. Optical absorption reveals an electronic phase transition near 5.4 GPa, and Raman results confirm its structural origin, which should be due to the paraelectric-incommensurate phase transition. It is noted that the Raman signal is still presented up to 30 GPa even if the sample turns into a superconductor above 15.3 GPa, suggesting a possible first-order structural phase transition between 15.3 and 30 GPa. This work shows that $Sn_2P_2Se_6$ is sensitive to external pressure and there are multiple phase transitions, which determine the FE insulating state, metallic state and even superconducting state, and will inspire the future comprehensive study on the correlation among FE/polar state, local structure and the superconductivity.